\documentclass[a4paper]{jpconf}
\usepackage{graphicx}
\usepackage{bm}

\newcommand{\thcrsi}{ThCr$_{2}$Si$_{2}$}

\newcommand{\rrusi}{$R$Ru$_{2}$Si$_{2}$}
\newcommand{\rerusi}[1]{#1Ru$_{2}$Si$_{2}$}
\newcommand{\ryrusix}{$R_{x}$Y$_{1-x}$Ru$_{2}$Si$_{2}$}
\newcommand{\dyyrusix}{Dy$_{x}$Y$_{1-x}$Ru$_{2}$Si$_{2}$}

\newcommand{\gdyrusix}{Gd$_{x}$Y$_{1-x}$Ru$_{2}$Si$_{2}$}
\newcommand{\dyyrusin}[2]{Dy$_{#1}$Y$_{#2}$Ru$_{2}$Si$_{2}$}
\newcommand{\tbyrusin}[2]{Tb$_{#1}$Y$_{#2}$Ru$_{2}$Si$_{2}$}
\newcommand{\gdyrusin}[2]{Gd$_{#1}$Y$_{#2}$Ru$_{2}$Si$_{2}$}
\newcommand{\femntiox}{Fe$_{x}$Mn$_{1-x}$TiO$_{3}$}

\newcommand{\Tg}{T_{\mbox{\scriptsize g}}}

\newcommand{\chinl}{\chi_{\mbox{\scriptsize nl}}}

\newcommand{\Ea}{E_{\mbox{\scriptsize a}}}

\newcommand{\gammaMF}{\gamma_{\mbox{\scriptsize MF}}}
\newcommand{\deltaMF}{\delta_{\mbox{\scriptsize MF}}}
\newcommand{\betaMF}{\beta_{\mbox{\scriptsize MF}}}

\newcommand{\znuMF}{(z\nu)_{\mbox{\scriptsize MF}}}

\begin{document}
\title{Critical Phenomena in Long-Range RKKY Ising Spin Glasses}

\author{Yoshikazu Tabata, Satoshi Kanada, Teruo Yamazaki$^{A}$, Takeshi Waki, Hiroyuki Nakamura
}

\address{
Department of Materials Science and Engineering, Kyoto University, Kyoto, Japan\\
$^{A}$The Institute for Soild State Physics,  University of Tokyo, Chiba, Japan\\
}

\ead{y.tabata@ht4.ecs.kyoto-u.ac.jp}

\begin{abstract}

We have investigated critical phenomena in spin glasses \ryrusix\ ($R$ $=$ Dy, Tb, Gd). These compounds, where the magnetic moments of rare-earth ions interact by the long-range Ruderman-Kittel-Kasuya-Yoshida (RKKY) interaction via conduction electrons, has uniaxial magnetic anisotropy. The separation of the zero-field-cooled and field-cooled magnetization was found only along the c-axis in all compounds, and hence, they are classified into the long-range Ising spin glass. The magnetic anisotropic energies in these compounds are different from each other in two orders of magnitude, from 330 K to 1.8 K, however, the critical exponents are similar. It clearly indicates a presence of the universality of the long-range RKKY Ising spin glasses. 

\end{abstract}

\section{Introduction}

Spin glass (SG) state is an archetype of ordered states in random magnets where the ferromagnetic (positive) and antiferromagnetic (negative) spin-spin interactions are randomly distributed. The strong frustration due to the random distribution of the interaction produces a complicated magnetic ordered state, where the spins are frozen with random orientations, showing striking glassy behaviors such as the long-time relaxation, aging phenomenon, memory effect, and chaos effect \cite{MemoryChaos,NatureGlassySG}. Although extensive investigations of the SG have been done since its discovery, many unresolved problems remains. One of them is whether the replica-symmetry-breaking (RSB) predicted by the mean-field theory \cite{ParisiRSB} occurs in real SG materials. 

Two alternative pictures of the SG state are theoretically predicted; the mean-field RSB picture and the droplet RS picture. According to theoretical studies of the mean-field model, the SG phase transition is a 2nd order phase transition associated with the RSB \cite{SKmodel,ParisiRSB}.  The RSB SG state has a complicated multi-valley structure of its free-energy \cite{MezardRSB} and is an essentially different state from RS states such as paramagnetic and (anti)ferromagnetic states, where simple one- and two-valley structures, respectively. Alternatively, the RS SG state is predicted by the droplet theory based on the short-range interaction model \cite{FisherHuseDroplet}. In the droplet picture, the SG state has a simple two-valley structure of its free energy as well as in an (anti)ferromagnetic state. The RSB and RS SG states can be distinguished by examination of stability of the SG phase under finite magnetic field. The RSB SG state is stable in the presence of finite magnetic field \cite{AT}, whereas the droplet RS SG state is unstable in any magnetic field \cite{FisherHuseDroplet}. 

Many numerical and experimental investigations were performed to examine the stability of the SG under magnetic field \cite{Numerical, NordbladFeMnTiO,JonssonFeMnTiO,TabataDYRS}.  The experiments of two different model SG materials, \femntiox\  \cite{NordbladFeMnTiO,JonssonFeMnTiO} and \dyyrusix\  \cite{TabataDYRS}, are worthy of attention. The former \femntiox\ and the latter \dyyrusix\ are model magnets of the Ising SGs with the short-range and long-range interactions, respectively. The short-range super-exchange interaction via oxigen-ion is dominant in \femntiox\ and the long-range Ruderman-Kittel-Kasuya-Yoshida (RKKY) interaction via conduction electrons is dominant in \dyyrusix . The experimental examinations on the stability of the SG under magnetic field were performed in both materials to verify critical divergence of a characteristic relaxation time $\tau (T,H)$ with finite SG transition temperature $\Tg(H)$. In \femntiox , $\tau (T,H)$ exhibits the critical divergence only at zero field \cite{NordbladFeMnTiO, JonssonFeMnTiO}. On the other hand, the critical divergence of $\tau (T,H)$ in \dyyrusix\ was found in both zero and finite fields \cite{TabataDYRS}. These results indicate that the RS and RSB SG states are realized in the short-range and long-range Ising SGs, respectively. Furthermore, the static and dynamic critical exponents of \femntiox\ and \dyyrusix\ are quite different from each other, i.e., $\gamma$ $=$ 4.0, $\beta$ $=$ 0.54, $\delta$ $=$ 8.4, and $z\nu$ $=$ 11 in \femntiox\ \cite{NordbladFeMnTiO,JonssonFeMnTiO,GunnarssonFeMnTiO} and $\gamma$ $=$ 1.1, $\beta$ $=$ 0.9, $\delta$ $=$ 1.7, and $z\nu$ $=$ 2.1 in \dyyrusix\ \cite{TabataDYRS}, indicating that these SGs belong to different classes each other and a spacial-range of interaction is relevant to SG phase transitions. The critical exponents of \dyyrusix\ are similar to those of the mean-field model, i.e., $\gammaMF$ $=$ 1, $\betaMF$ $=$ 1, $\deltaMF$ $=$ 2, and $\znuMF$ $=$ 2 \cite{SGreview}. 
 
In this article, we have reported on static critical phenomena in spin glasses \ryrusix\ ($R$ $=$ Tb, Gd) in addition to \dyyrusix . The parent compounds of these SGs, \rerusi{Dy}, \rerusi{Tb}, and \rerusi{Gd}, are isostructural intermetallic compounds, the tetragonal \thcrsi\ structure,  with paramagnetic-antiferromagnetic transitions of 29 K, 53 K, and 47 K, respectively \cite{RRuSi}. In \rerusi{Tb} and \rerusi{Gd}, the long-range RKKY interaction is dominant and spin glass (SG) phase also appears by substitution of non-magnetic Y for $R$, as well as in \rerusi{Dy}. Although magnetic anisotropic energies in these compounds are different in two orders of magnitude , from 330 K to 1.8 K, the magnetic anisotropy is the same, an Ising-like, and static critical exponents, $\gamma$ and $\delta$, 
are similar to each other. It clearly indicates a presence of the universality of the long-range RKKY Ising SGs. 
  
\section{Experimental details}

The single crystalline samples of \ryrusix\ ($R$ $=$ Dy, Tb, Gd) were grown by the Czhochralski method with a tetra-arc furnace. The concentrations of $R$ were determined by comparing the saturated magnetizations of the diluted compounds along the magnetic easy c-axis with those of the pure compounds \rrusi . Ac and dc magnetizations were measured with the SQUID magnetometer MPMS (Quantum Design) equipped in the Research Center for Low Temperature and Material Sciences in Kyoto University. The dc magnetization was measured 10 minutes after stabilization at a certain temperature and a certain field to avoid the effect of very slow dynamics and to obtain the thermodynamic equilibrium magnetizations near the SG transition temperatures. The ac measurement of \gdyrusix\ was performed with an ac field of 1 Oe and a frequency of 0.5 Hz. We use thin plate-shaped samples with weights of about 7.0 mg and sizes of $5 \, \times 1 \, \times 0.2$ mm$^{3}$ for the measurements to inhibit the effects of the diamagnetic field and the eddy current. 

To examine the static critical behaviors, we extracted the second nonlinear susceptibility $\chi_{2}(T)$  from the dc magnetic field ($H$) dependences of the dc magnetization $M(H,T)$ and the real part of the ac susceptibility $\chi'(H,T)$.  From a dc magnetization measurement, $\chi_{2}(T)$ can be obtained by fitting $M(H,T)/H$ at each temperature as a function of $H^{2}$ in the form of 
\begin{equation}
M(H,T)/H \, = \, \chi_{0} (T) + \chi_{2} (T) H^{2} + \cdots . 
\label{MoverHvsH2}
\end{equation}
$\chi_{2} (T)$ can also be obtained from an ac susceptibility by fitting $\chi' (H,T)$ as a function of $H^{2}$ in the form of 
\begin{equation}
\chi' (H,T) \, = \, \chi'_{0} (T) + 3 \chi'_{2} (T) H^{2} + \cdots , 
\label{ChiacvsH2}
\end{equation}
because $\chi' (H,T)$ with small ac field represents a differential magnetization $dM/dH$.  When an ac-frequency $\omega$ is so low that we can recognized the measurement to be in the dc-limit, $\chi'_{2}(T)$ obtained from the ac measurement is equivalent to $\chi_{2}(T)$ obtained from the dc measurement. In the present work, we obtained $\chi_{2}(T)$'s of \dyyrusin{0.103}{0.897} and \tbyrusin{0.065}{0.935} from the dc measurements because they show strong frequency dependence of the ac susceptibilities even very low frequency of 0.1 Hz. \gdyrusin{0.057}{0.943} hardly shows the frequency dependence and is within the dc limit below 1 Hz, and hence, we obtained its $\chi_{2}(T)$ from the ac measurement with $\omega$ $=$ 0.5 Hz. The field-depedent nonlinear susceptibility $\chinl(H,T)$ $\equiv$ $\chi_{0}(T) - M(H,T)/H$ was extracted by using the linear susceptibility $\chi_{0}(T)$ obtained from above-mentioned procedures. 

\section{Results and discussions}

\begin{figure}[ht]
\begin{center}
 \includegraphics[clip,width=16cm]{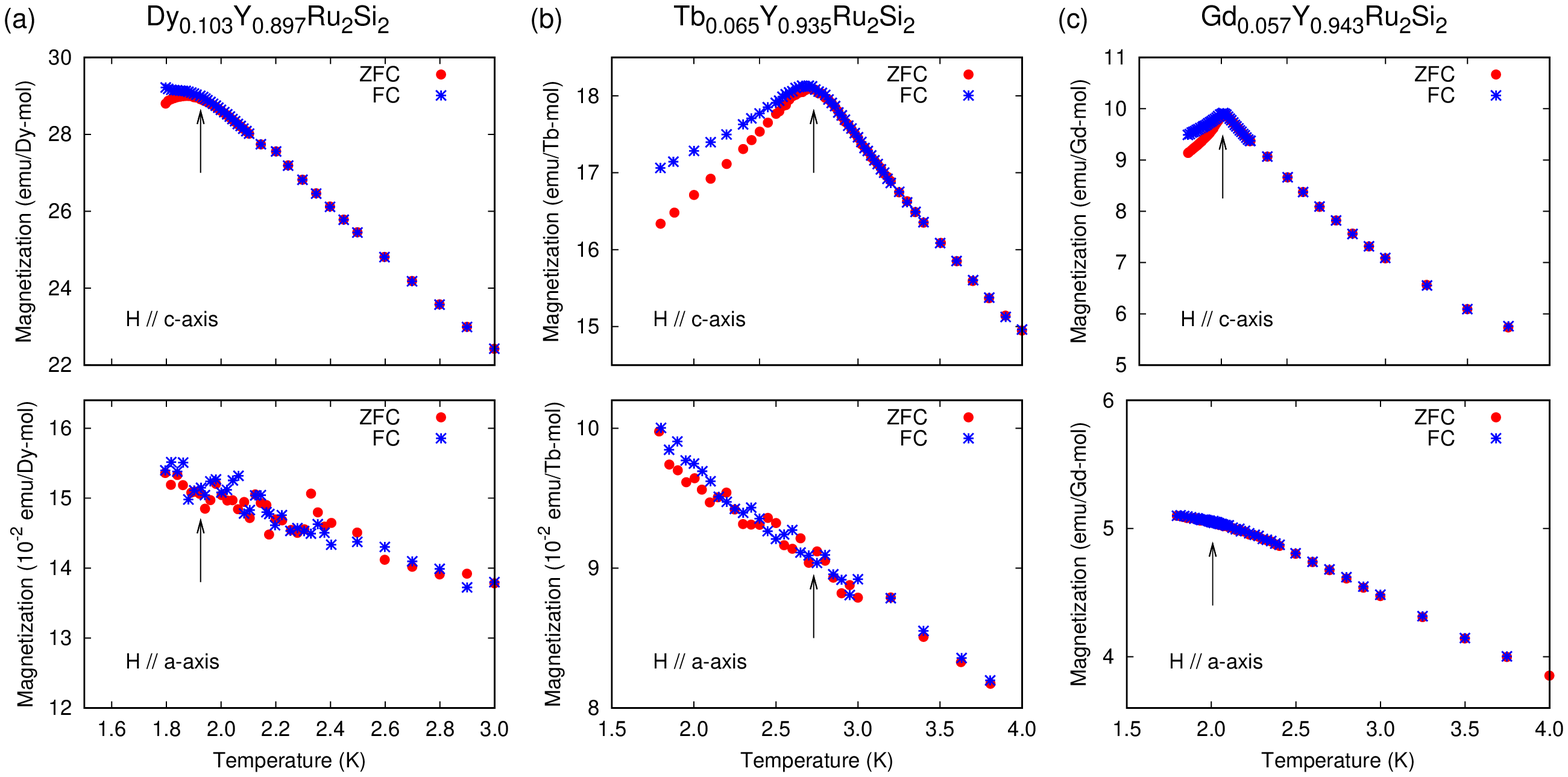}
\end{center}
\caption{\label{fig1} $T$ dependences of dc magnetizations of (a) \dyyrusin{0.103}{0.897}, (b) \tbyrusin{0.065}{0.935}, and (c) \gdyrusin{0.057}{0.943}. Upper and lower figures are the data with magnetic field along the c-axis and a-axis, respectively. Arrows indicate the SG transition temperatures.}
\end{figure}

Figure \ref{fig1} shows $T$ dependences of dc magnetizations of (a) \dyyrusin{0.103}{0.897}, (b) \tbyrusin{0.065}{0.935} and (c) \gdyrusin{0.057}{0.943} along the c- and a-axes with the applied magnetic field of 2 Oe. All compounds exhibit distinct separations of the zero-field-cooled (ZFC) and field-cooled (FC) magnetizations along the c-axis around 1.9 K (Dy), 2.7 K (Tb), and 2.0 K (Gd). In contrast, no separations of the ZFC and FC magnetizations along the a-axis are found. These results imply that only the c-component of magnetic moments of the magnetic Dy$^{3+}$-, Tb$^{3+}$-, and Gd$^{3+}$-ions are randomly frozen and the a-component dynamically fluctuates even at the lowest temperature of measurements. These component-separated freezing behaviors, which was theoretically predicted in the anisotropic SG based on the mean-field model \cite{AnisoSG}, indicate that these compounds should be categorized into the Ising SG with the long-range RKKY interaction. 

\begin{figure}[ht]
\begin{center}
 \includegraphics[clip,width=16cm]{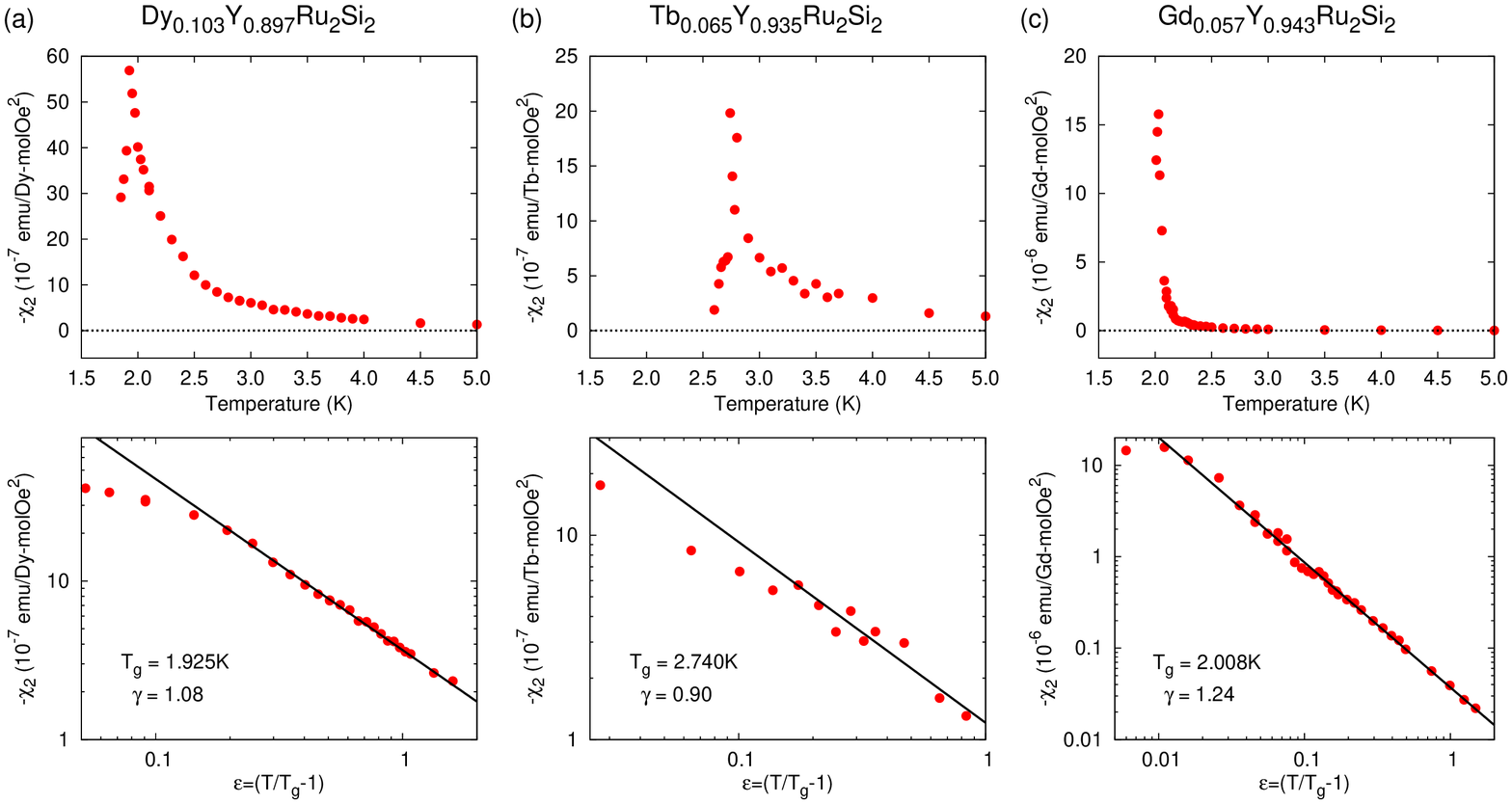}
\end{center}
\caption{\label{fig2} $T$ dependences of second nonlinear susceptibilities of (a) \dyyrusin{0.103}{0.897}, (b) \tbyrusin{0.065}{0.935}, and (c) \gdyrusin{0.057}{0.943}. Lower figures are the log-log plots of $-\chi_{2}$ vs $\varepsilon$ ($\equiv$ $T/\Tg -1$).}
\end{figure}

Figure \ref{fig2} shows $T$ dependences of second nonlinear susceptibilities $\chi_{2}(T)$'s, corresponding to the order-parameter susceptibility of the SG,  (a) \dyyrusin{0.103}{0.897}, (b) \tbyrusin{0.065}{0.935} and (c) \gdyrusin{0.057}{0.943} along the c-axis. In all compounds, clear divergent behavior of $\chi_{2}(T)$ are found. From the log-log plots of $-\chi_{2}$ vs $\varepsilon$ ($\equiv$ $T/\Tg -1$) shown in the lower figures of Fig. \ref{fig2},  we estimated the critical exponent $\gamma$ of $-\chi_{2}(T)$ $\propto$ $\varepsilon ^{-\gamma}$. The best plots showing linear relations between $\log (-\chi_{2})$ and $\log \varepsilon$ with $\Tg$ $=$ 1.925 K (Dy), 2.740 K (Tb), and 2.008 K (Gd). The estimated $\gamma$'s are 1.08, 0.90, and 1.24 in \dyyrusin{0.103}{0.897}, \tbyrusin{0.065}{0.935} and \gdyrusin{0.057}{0.943} respectively. These values of $\gamma$ are similar to each other and are approximately the value of the mean-field model $\gammaMF$ $=$ 1. 

\begin{figure}[ht]
\begin{center}
 \includegraphics[clip,width=16cm]{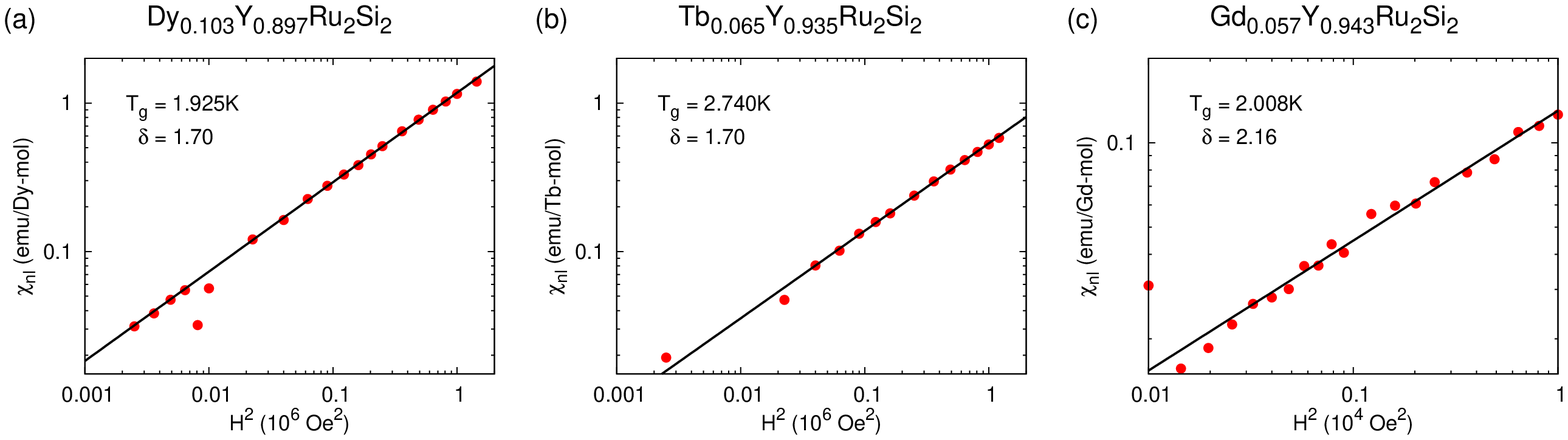}
\end{center}
\caption{\label{fig3} Log-log plots of $H^{2}$ dependences of field-dependent nonlinear susceptibilities of (a) \dyyrusin{0.103}{0.897}, (b) \tbyrusin{0.065}{0.935}, and (c) \gdyrusin{0.057}{0.943} at each $\Tg$.}
\end{figure}

Next, we show $H^{2}$ dependences of field-dependent nonlinear susceptibilities at $\Tg$ $\chinl (H,\Tg)$, corresponding to the SG order-parameter,  in Figs \ref{fig3} (a) \dyyrusin{0.103}{0.897}, (b) \tbyrusin{0.065}{0.935}, and (c) \gdyrusin{0.057}{0.943} along the c-axis. In the figures, $\chinl$ vs. $H^{2}$ are plotted in double logarithmic scales. From these plots, the critical exponent $\delta$ of $\chinl$ $\propto$ $H^{2/\delta}$ is obtained as $\delta$ $=$ 1.70 (Dy), 1.70 (Tb), and 2.16 (Gd). These values of $\delta$ are similar to each other and are approximately the value of the mean-field model $\deltaMF$ $=$ 2, as well as $\gamma$. 

The similarity of the static critical exponents $\gamma$ and $\delta$ indicates that the SG phase transitions of \ryrusix\  should belong to the same universality class, which is different from those of not only \femntiox\ \cite{GunnarssonFeMnTiO} but also so-called canonical SGs \cite{LevyAgMn,TaniguchiAuFe}. The anisotropic behaviors shown in Fig. \ref{fig1} clearly indicate the Ising nature of \ryrusix . We roughly estimated magnetic anisotropic energies $\Ea$'s from the differences between magnetization processes along the c- and a-axes; $\Ea$ $=$ 250, 330, and 1.8 K/$R^{3+}$-ion in  \dyyrusin{0.103}{0.897}, \tbyrusin{0.065}{0.935} and \gdyrusin{0.057}{0.943} respectively. Moderate magnitude of $\Ea$ ($\approx$ $\Tg$), which is enough large anisotropy to show the Ising nature of SG phase transition according to the theory \cite{AnisoSG}, is found even in Gd-compound with zero orbital angular momentum ($L$ $=$ 0). It should be noted that the critical exponents in \ryrusix\ are almost identical in spite of the large difference of $\Ea$, and are similar to those of the mean-field model. 

Here, we discuss on the validity of our static scaling analyses. The cusp-like anomalies of the ZFC magnetization of the \dyyrusin{0.103}{0.897} and \tbyrusin{0.065}{0.935} are rather broad, as shown in Figs \ref{fig1} (a) and (b). It should be a dynamical effect because very slow dynamics is found in both compounds \cite{TabataDYRS}. The slow dynamics in the Dy- and Tb-compounds result from the strong Ising anisotropy and the consequent high energy barrier of the spin flipping. Due to this dynamical effect, it is very difficult to observe the static critical properties very close to $\Tg$ in the Dy- and Tb-compounds.  Thus, the regions of the measurements in these compounds shown in Figs. \ref{fig2} and \ref{fig3} are rather far from $\Tg$, as comparing with that in the earlier work of the canonical SG AgMn \cite{LevyAgMn}. The analysis of the data in rather high temperature and high field may lead wrong critical exponents. On the other hand, the magnetic anisotropy of the \gdyrusin{0.057}{0.943} is weak and the effect of the slow dynamics is negligible above $\Tg$, and hence, we can approach very close $\Tg$, as shown in Figs. \ref{fig2} (c) and \ref{fig3} (c). Nevertheless, we obtained the critical exponents, $\gamma$ $=$ 1.24 and $\delta$ $=$ 2.16, which are roughly identical to the exponents of the Dy- and Tb-compounds and are much closer to the mean-field exponents than to those of the short-range Ising SG or the canonical SG \cite{noteLRSG}. Consequently, we concluded that \ryrusix , the long-range RKKY Ising SG, belongs to the mean-field universality class.

\section{Conclusion}

We examined the static critical phenomena of the SG phase transition in \ryrusix\ ($R$ $=$ Dy, Tb, Gd), which are categorized as the long-range RKKY Ising SG, and found almost identical critical exponents $\gamma$ $\sim$ 1 and $\delta$ $\sim$ 2. These values are similar to the mean-field values $\gammaMF$ $=$ 1 and $\deltaMF$ $=$ 2. It indicates a presence of the universality of the long-range RKKY Ising spin glasses, being the mean-field universality class.

\section{Acknowledgements}

This study was supported by a Grand-in-Aid for Scientific Research on Priority Areas "Novel States of Matter Induced by Frustration" (19052003), a Grand-in-Aid for the Global COE Program "International Center for Integrated Research and Advanced Education in Materials Science" from the Ministry of Education, Cluture, Sports, Science and Technology of Japan. 


\end{document}